\documentclass[aps,prl,reprint,groupedaddress,showpacs,showkeys]%
{revtex4-1}
\usepackage{graphicx}
\usepackage{amsmath}
\usepackage{amssymb}
\usepackage{psfrag}

\def\w{\omega}

\def\e{\varepsilon}

\newcommand{\vp}{\varphi}

\renewcommand{\geq}{\geqslant}
\renewcommand{\le}{\leqslant}
\renewcommand{\ge}{\geqslant}

\begin{document}
\title{Solitary state at the edge of synchrony in ensembles with 
attractive and repulsive interaction}

 \author{Yuri Maistrenko$^{1,2}$, Bogdan Penkovsky$^{3,4}$,
Michael Rosenblum$^{1}$}
 \affiliation{
$^1$Department of Physics and Astronomy, University of Potsdam,
  Karl-Liebknecht-Str. 24/25, D-14476 Potsdam-Golm, Germany\\
$^2$Inst. of Mathematics and National Centre for Medical and Biotechnical
Research, National Academy of Sci. of Ukraine, Kiev, Ukraine\\
$^3$National University of ``Kyiv-Mohyla Academy'', Kiev, Ukraine \\
$^4$FEMTO-ST/Optics Department, University of Franche-Comt\'{e}, 16 Route de Gray, 25030 Besan\c{c}on Cedex, France
}

\begin{abstract}
We discuss the desynchronization transition in networks of globally coupled identical oscillators with attractive 
and repulsive interactions. 
We show that, if  attractive and repulsive groups act in antiphase or close to that, a \textit{solitary state} 
emerges with a single repulsive oscillator split up from the others fully synchronized.  With further increase of the 
repulsing strength, the synchronized cluster becomes \textit{fuzzy} and the dynamics is given by a variety of stationary 
states with zero common forcing.  Intriguingly, solitary states represent the natural link between coherence and 
incoherence. The phenomenon is described analytically for phase oscillators with sine coupling and demonstrated 
numerically for more general amplitude models.
\end{abstract}

\date{\today}

\maketitle
Mean field approximation, or global coupling, is widely used in description 
of oscillator networks with high degree of connectivity. 
In case of weak interactions, the theoretical analysis of the dynamics is 
typically performed with the help of phase approximation 
\cite{Kuramoto-75,*Kuramoto-84,Daido-92a,*Daido-93a}, 
most frequently with the use of the analytically solvable Kuramoto-Sakaguchi model
\cite{Kuramoto-84,Sakaguchi-Kuramoto-86,Watanabe-Strogatz-93,%
*Watanabe-Strogatz-94,Ott-Antonsen-08}. 
A topic of recent interest is investigation of interaction of 
several globally coupled ensembles
\cite{Okuda-Kuramoto-91,*Montbrio_etal-04,*Abrams-Mirollo-Strogatz-Wiley-08,%
*Sheeba_et_al-08,*Martens-10,*Martens-10a,%
*Skardal-Restrepo-12,Pikovsky-Rosenblum-08}, 
in particular with attracting (positive) and repulsive (negative) 
couplings \cite{Zanette-05,Hong-Strogatz-11a,*Hong-Strogatz-11b}. These studies
are partially motivated by the problems of neuroscience where many highly connected groups of 
neurons interact via  excitatory and inhibitory connections 
\cite{Wilson-72,*vanVreeswijk-96,*Kopell-2003,*Peyrachea-2012,*Ledoux-Brunel-2011,*Carcea-Froemke-2013}. 

One of the intriguing effects in ensembles of globally
coupled identical oscillators is clustering (see e.g. \cite{Pikovsky-Popovych-Maisternko-2001,Ashwin-2007} 
and references therein). 
It appears that 
randomly chosen initial states in the course of evolution can eventually become identical, 
and the final configuration consists of 
clusters of identically equal states.
In this Letter, we discuss the formation of clusters and desynchronization transition in 
finite-size ensembles of identical oscillators with attractive and 
repulsive coupling and demonstrate a novel scenario: when the repulsion
starts to prevail over the attraction, 
a solitary oscillator leaves the synchronous cluster creating so-called \textit{solitary state}. 
With further increase of the repulsion, solitary state loses its stability.
More and more oscillators leave the synchronous group, which becomes a \textit{fuzzy cluster}.
Our aim is to describe this scenario  for $M$-group Kuramoto-Sakaguchi model 
\begin{equation*}
 \dot\theta^{\sigma}_i = \omega+\sum_{\sigma'=1}^{M}\frac{K_{\sigma\sigma'}}{N}\sum_{j=1}^{N_{\sigma'}}\sin(\theta^{\sigma'}_j-\theta^{\sigma}_i+\alpha_{\sigma\sigma'}),  
\label{eqmodM}
\end{equation*}
where $\theta^{\sigma}_i$ is the phase of the $i$-th oscillator in group $\sigma$, 
%$i=1,\ldots,N_\sigma$,  
$N_{\sigma}$ is the number of oscillators in the group, and 
$N=N_1+...+N_M$.  Elements of the $M\times{M}$ matrices $K_{\sigma\sigma'}$ and $\alpha_{\sigma\sigma'}$ 
represent the coupling strength and the phase shift of each oscillator in group $\sigma'$ acting on each 
oscillator in group $\sigma$. By transformation to a co-rotating coordinate frame we can put $\w=0$.

We introduce the effect starting with the two-group model ($M=2$), assuming $K_{\sigma\sigma'}=K_{\sigma'}$, $\alpha_{\sigma\sigma'}=\alpha_{\sigma'}$ for all $\sigma,\sigma'=1,2$, i.e., that the coupling strengths and the 
phase shifts are determined by the acting group only:
\begin{equation}
\dot\theta^{\sigma}_i = \sum_{\sigma'=1}^{2}\frac{K_{\sigma'}}{N}\sum_{j=1}^{N_{\sigma'}}\sin(\theta^{\sigma'}_j-\theta^{\sigma}_i+\alpha_{\sigma'})\;.  
\label{eqmod2}
\end{equation}
Furthermore, we suppose $K_1>0$, $K_2<0$ and $-\pi/2<\alpha_{1},\alpha_{2}<\pi/2$ such that the first group acts
attractively on all oscillators in  the network and the second group acts repulsively, 
cf.~\cite{Anderson_et_al-12}.
Such coupling configuration is a prototype of  neuronal networks with excitatory ($K_1>0$)  and  
inhibitory ($K_2<0$) neurons 
\cite{Wilson-72,*vanVreeswijk-96,*Kopell-2003,*Peyrachea-2012,*Ledoux-Brunel-2011,*Carcea-Froemke-2013}. 

By re-normalizing the time, $t\to K_1t$, we write the coupling coefficients  as $K_1=1$, $K_2=-(1+\e)$,
 where the new coupling parameter 
\begin{equation}
\e=-(1+K_2/K_1)
\end{equation}
quantifies the \textit{excess of the repulsion over the attraction}. 
Introducing the complex order parameter for both groups 
$Z_{\sigma}=N_{\sigma}^{-1} \sum_{j=1}^{N_{\sigma}}e^{i\theta^{\sigma}_j}=\rho_{\sigma}e^{i\Theta_{\sigma}}$
and re-labeling the phases as 
$\theta_j=\theta^{\sigma}_i$, where $j=1,\ldots,N$, $j=(\sigma-1)N_1+i$, 
%(such that $j=i$ for $\sigma=1$  and $j=N_1+i$ for $\sigma=2$), 
we bring the system to the form $ \dot\theta_j = h\sin(\Phi-\theta_j)$ 
with the common forcing
\begin{equation}
H=he^{i\Phi}= \frac{N_1}{N}e^{i\alpha}Z_1  - \frac{N_2}{N}( 1+\e)e^{i\beta}Z_2\;,
\label{hforce}
\end{equation}
where, for convenience, we rename $\alpha=\alpha_1$ and $\beta=\alpha_2$.

We emphasize, that, although the oscillators of two groups contribute differently to $H$, 
they evolve under the common forcing and therefore the whole population 
is effectively three-dimensional: 
it can be described by three Watanabe-Strogatz (WS) equations for collective variables 
$\kappa,\Psi,\Gamma$
\cite{Watanabe-Strogatz-93,*Watanabe-Strogatz-94,Pikovsky-Rosenblum-08,Pikovsky-Rosenblum-11}.
These feature distinguishes our model from the ``conformists and contrarians''
model by Hong and Strogatz (HS) 
\cite{Hong-Strogatz-11a,*Hong-Strogatz-11b}.
%\footnote{
%In the HS model 
%all oscillators contribute positively to the mean field and differ by the way they %react 
%to it. The HS model is described by 
%two coupled three-dimensional systems of the WS equations, i.e. is six-dimensional.}.

The WS equations contain $N$ constants of motion $\chi_j$, 
determined from initial conditions; $\chi_j$ obey three additional constrains. 
Notice that $0\le\kappa\le 1$ while $\Psi,\Gamma$, and $\chi_j$ are angles.
The original phases $\theta_j$ are restored by the  transformation
$e^{i\theta_j}=e^{i\Gamma}\left(\kappa+e^{i(\chi_j-\Psi)}\right )\left(\kappa e^{i(\chi_j-\Psi)}+1\right)^{-1}$.
If the system evolves to a state with $\kappa=1$, then all initially different phases become 
identical (one-cluster state). 
Exceptional is the case when $\kappa=1$ and $\chi_j-\Psi=\pi$ for some $j=n$ 
\footnote{A. Pikovsky, private communication.}; 
then $\theta_n$ may differ from all other phases.
Such \textit{solitary states}, when all the phases but one are identical, are of our 
main interest here.
Below we show that stable solitary states naturally appear in our model in course of the 
desynchronization transition.
Notice that other clustered states, except for  fully synchronous and solitary ones, 
contradict the WS theory. 
Instead, the model exhibits a variety of neutrally stable \textit{fuzzy clusters}, 
where some number of oscillators are split up from the others ``almost'' synchronized. 

To describe the desynchronization transition we first check that the fully synchronous state 
$\theta_j\equiv\vp$ of model (\ref{eqmod2}) is stable for 
\begin{equation}
\e<\e_{cr}=\frac{N_1\cos\alpha}{N_2\cos\beta}-1\;.
\label{ecrit}
\end{equation}
For $\e>\e_{cr}$, we look for a solitary state $\theta_1=\ldots=\theta_{N-1}\equiv\vp$, $\theta_N\equiv\psi$, i.e. when 
one repulsive unit splits up from all others 
\footnote{Since all oscillators are identical we can without a loss of generality take that the solitary 
unit has index $N$.}.
Dynamics of this state are given by two equations which can be easily obtained by 
direct substitution of $\vp$ and $\psi$ into Eq.~(\ref{eqmod2}):
\begin{equation}
\begin{aligned}
\dot\vp &=-\frac{1+\e}{N}[\sin(\eta+\beta)+
(N_2-1)\sin\beta]+\frac{N_1}{N}\sin\alpha,\\
\dot\psi &= \frac{1+\e}{N}[(N_2-1)\sin(\eta-\beta)-
\sin\beta]-\frac{N_1}{N}\sin(\eta-\alpha)\;,
\end{aligned}
\label{eqlam1}
\end{equation}
where we denote $\eta=\psi-\vp$. After straightforward manipulations, this system can be reduced 
to a scalar equation for the phase difference $\eta$:
\begin{equation}
\dot\eta = A[\sin(\eta-\eta^{\ast})+\sin\eta^{\ast}]\;.
\label{eqlam5}
\end{equation}
Here $A\geq0$,  and $\eta^*$ is expressed, using $p=N_1/N$, as
\begin{equation}
\eta^{\ast}=\arctan\frac{(1-p-2N^{-1})(1+\e)\sin\beta - p\sin\alpha}
{(1-p)(1+\e)\cos\beta - p\cos\alpha}\;.
\label{eqlam101}
\end{equation}
Equation~(\ref{eqlam5}) has two equilibria $\eta_{syn}=0$ and $\eta_{sol}=2\eta^{\ast}+\pi$ which 
describe, respectively,  the full synchrony ($\psi=\vp$) and the solitary state ($\psi=\vp+\eta_{sol}$) in the original 
model (\ref{eqmod2}). It can be easily checked 
\footnote{Stability/instability of the states $\eta_{syn}=0$ and $\eta_{sol}$ is determined by the 
sign of $\cos\theta=(1-p)(1+\e)\cos\beta - p\cos\alpha$.}
that these states exchange their stability exactly 
at $\e_{cr}$. Hence, the solitary state is stable for all $\e>\e_{cr}$ within the two-cluster manifold 
$(\vp, \psi)$.
To complete the stability analysis of the solitary state we have to examine in the phase space the directions, transversal to  
the manifold $(\vp,\psi)$, i.e. to quantify the stability of the main synchronized cluster of $N-1$ elements. 
For this goal we write the Jacobian for the system~(\ref{eqlam1}) 
at solitary state $\psi=\vp+\eta_{sol}$. Due to the matrix symmetry we find that Jacobian has $N-2$ 
equal eigenvalues 
\footnote{Indeed, substituting $\lambda=\lambda_\bot$ in the matrix $J-{\lambda}E$ 
immediately yields $N-1$ identical rows.} which are, actually, transversal LE of the solitary state, all equal:
\begin{equation}
\lambda_\bot= - p\cos\alpha+(1-p-\frac{1}{N})(1+\e)\cos\beta-\frac{1+\e}{N}\cos(2\eta^{\ast}+\beta).
\label{transLE}
\end{equation}

\begin{figure}[ht!]
\centerline{\includegraphics[width=0.5\textwidth]{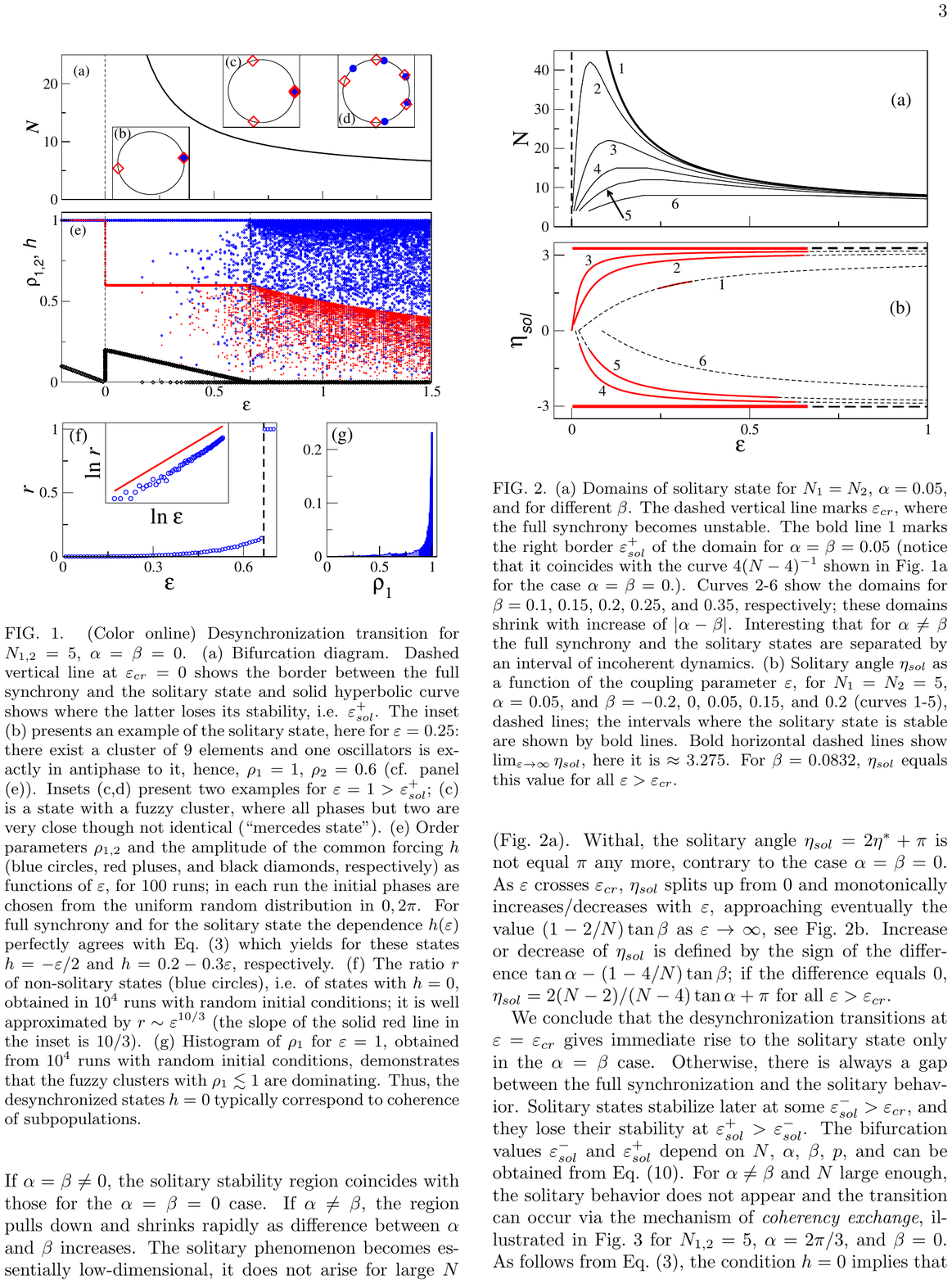}}
%\centerline{\includegraphics[width=0.45\textwidth]{bifdiag.pdf}}
%\centerline{\includegraphics[width=0.45\textwidth]{stat_hist.pdf}}
\caption{(Color online) Desynchronization transition for $N_{1,2}=5$, $\alpha=\beta=0$.
(a) Bifurcation diagram. Dashed vertical line at $\e_{cr}=0$ shows the border between the full 
synchrony and the solitary state and 
solid hyperbolic curve shows where the latter loses its stability, i.e. $\e^+_{sol}$. 
The inset (b) presents an example of the solitary state, here for $\e=0.25$: 
there exist a cluster of 9 elements and one oscillators is exactly in antiphase to it, 
hence, $\rho_1=1$, $\rho_2=0.6$ (cf. panel (e)).
Insets (c,d) present two examples for $\e=1>\e^+_{sol}$; (c) is a state with a fuzzy cluster, 
where all phases but two are very close though not identical (``mercedes state'').
(e) Order parameters $\rho_{1,2}$ and the amplitude of the common forcing $h$
(blue circles, red pluses, and black diamonds, respectively) as functions of $\e$,
for $100$ runs; in each run the initial phases are chosen from the uniform random distribution 
in $0,2\pi$.  
For full synchrony and for the solitary state the dependence $h(\e)$ perfectly agrees with
Eq.~(\ref{hforce}) which yields for these states $h=-\e/2$ and $h=0.2-0.3\e$, respectively.
(f) The ratio $r$ of non-solitary states (blue circles), i.e. of states with $h=0$, 
obtained in $10^4$ runs with random initial conditions; 
it is well approximated by $r\sim \e^{10/3}$ (the slope of the solid red line in the inset is $10/3$).
(g) Histogram of $\rho_1$ for $\e=1$, obtained from $10^4$ runs with random initial conditions, 
demonstrates that the fuzzy clusters with $\rho_1\lesssim 1$ are dominating. 
Thus, the desynchronized states $h=0$ typically correspond to coherence of subpopulations.
}
\label{bifdiag}
\end{figure}

For interpretation of the results we concentrate first on the simplest nontrivial case $N_1=N_2=N/2$  and 
$\alpha=\beta=0$.  Then $\eta_{sol}=\pi$, i.e., the solitary oscillator stays strictly in 
antiphase to all others, and Eqs.~(\ref{ecrit},\ref{transLE}) yield the stability domain 
of this state 
\begin{equation}
0<\e<\e_{sol}^{+}=4(N-4)^{-1} \;,
\label{stabregion}
\end{equation}
see Fig.~\ref{bifdiag}a. 
It follows that solitary state which is stationary in this case exists for arbitrarily large network size $N$, 
however, the $\e$-width of the stability domain shrinks to zero as $N\to\infty$.
Our numerical studies reveal, however, that basin of attraction for the solitary state can be not of full measure.
Indeed, solutions with $h=0$ (see Eq.~(\ref{hforce})), which are
also stationary, coexist with the solitary state in the stability domain (\ref{stabregion}). As it is illustrated in
Fig.~\ref{bifdiag}e, immediately after the transition at $\e_{cr}$  solitary states appear with 
probability one; soon after, the $h=0$ states arise with non-zero probability which grows 
$\sim \e^{10/3}$ as $\e$ approaches $\e_{sol}^{+}$.
For $\e>\e_{sol}^{+}$ solitary state does not exist any more.
%\footnote{
%As $\e$ crosses the bifurcation value $\e_{cr}^{+}=4/(N-4)$, all $N-2$ transverse Lyapunov exponents of the solitary 
%state simultaneously become positive, violating its stability. The state still exists and
%is stable inside the corresponding two-dimensional clustered manifold $(\vp,\psi)$ but it is already 
%vastly unstable transversely and any small perturbations transform the main $N-1$-dimensional 
%synchronized cluster into a fuzzy set}.
All stationary states fulfill the condition $h=0$; then Eq.~(\ref{hforce}) yields
$\rho_1=\rho_2(1+\e)$. It turns out, that most likely are the states 
when the attractive units form a fuzzy cluster with $\rho_1\lesssim 1$, and, respectively, 
$\rho_2\approx (1+\e)^{-1}$. This is illustrated in Fig.~\ref{bifdiag}e-g, where we present the results 
for numerical analysis with random initial conditions
\footnote{
An example of a fuzzy cluster for $\e>\e_{cr}^{+}$ is given in Fig.~\ref{bifdiag}c, cf. \cite{Zanette-05}.
It resembles the ``mercedes'' state, 
where two oscillators from  the repulsive group have phase shift
$\pm\arccos\left (1-\frac{\e N}{4(1+\e)}\right )$ with respect to $N-2$ fully synchronous units.  
However, as we have proved analytically,
the latter state is not attractive: it is stable inside its three-dimensional clustered manifold for
$4/{N-4}<\e<8/{N-8}$, but is only neutrally stable transversally. 
Further increase of $\e$ gives birth to analogous states with three and more oscillators split up 
from the main synchronized group.  Similarly, they are neutrally stable.}.

\begin{figure}[ht!]
%\centerline{\includegraphics[width=0.45\textwidth]{fig2.pdf}}
\centerline{\includegraphics[width=0.5\textwidth]{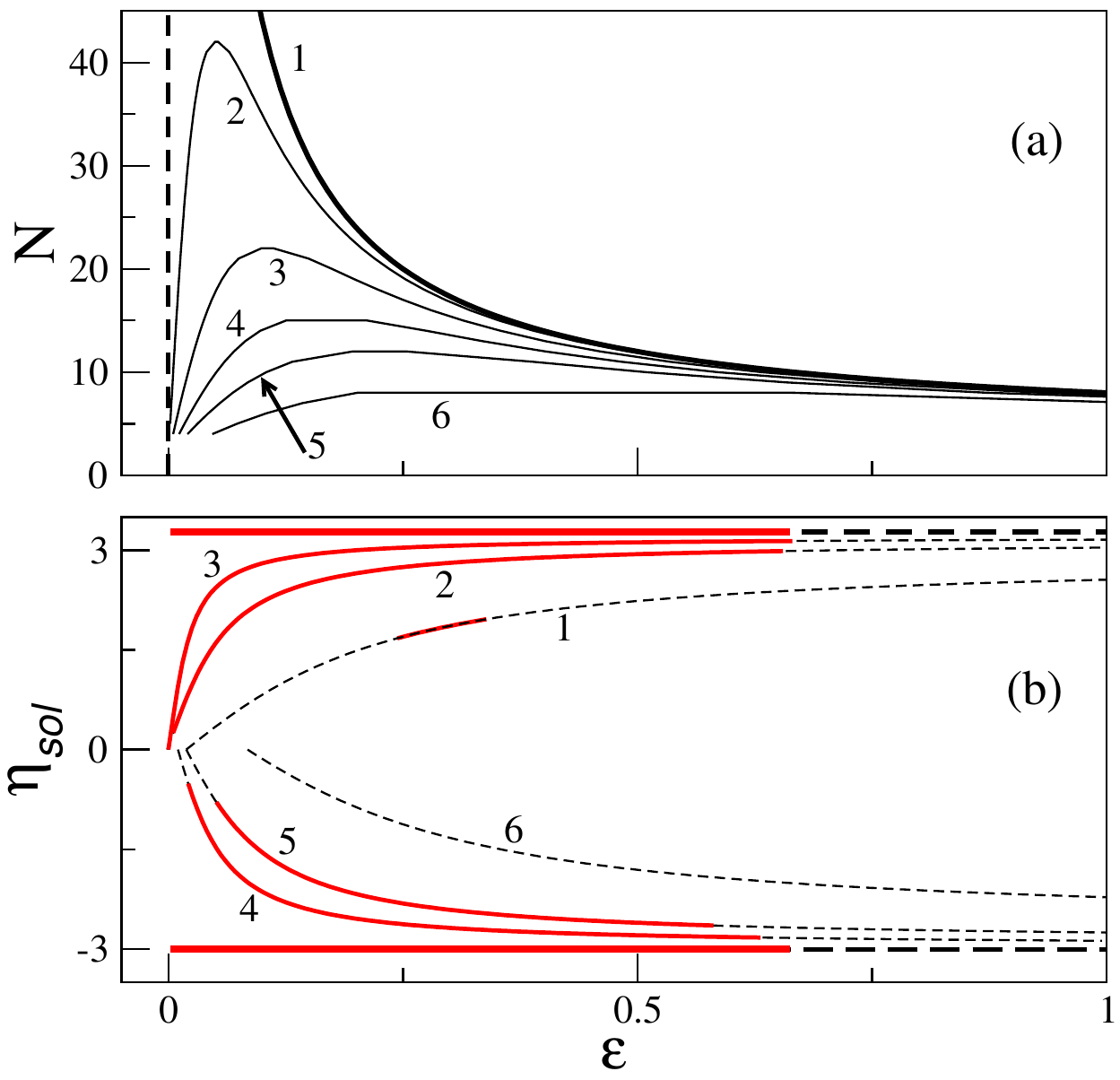}}
\caption{(a) Domains of solitary state for $N_1=N_2$,  $\alpha=0.05$, and for different $\beta$.  
The dashed vertical line marks $\e_{cr}$, where the full synchrony becomes unstable. The bold line 1
marks the right border $\e_{sol}^+$ of the domain for $\alpha=\beta=0.05$ (notice that it coincides with 
the curve $4(N-4)^{-1}$ shown in Fig.~\ref{bifdiag}a for the case $\alpha=\beta=0$.). 
Curves 2-6 show the domains for  $\beta=0.1$, $0.15$, $0.2$, $0.25$, and $0.35$,  respectively;
these domains shrink with increase of $|\alpha - \beta|$.  Interesting that for $\alpha\ne\beta$
the full synchrony and the solitary states are separated by an interval of incoherent dynamics.
(b) Solitary angle $\eta_{sol}$ as a function of the coupling parameter $\e$, for $N_1=N_2=5$, $\alpha=0.05$, and
$\beta=-0.2$, $0$, $0.05$, $0.15$, and $0.2$ (curves 1-5),  dashed lines; the intervals where the solitary state 
is stable are shown by bold lines. Bold horizontal dashed lines show  $\lim_{\e\to\infty}\eta_{sol}$, here it is 
$\approx 3.275$. For $\beta=0.0832$, $\eta_{sol}$ equals this value for all $\e>\e_{cr}$.
}
\label{diag2}
\end{figure}

Now, we consider the case $\alpha\ne 0$, $\beta\ne 0$, see Fig.~\ref{diag2}. 
The solitary state is not stationary any more (as follows from Eq.~(\ref{eqlam1}), 
all units rotate with a constant velocity), 
and its stability domain is obtained from Eqs.~(\ref{eqlam101},\ref{transLE}) as:
\begin{equation}
N<\frac{\cos\beta+\cos(2\eta^{\ast}+\beta)}{(1-p)\cos\beta-p(1+\e)^{-1}\cos\alpha}\;.
\label{eqlam103}
\end{equation}
If $\alpha=\beta\neq0$, the solitary stability region coincides with those for the $\alpha=\beta=0$ case.
If $\alpha\neq\beta$, the region pulls down and shrinks rapidly as difference between $\alpha$ and $\beta$ increases. 
The solitary phenomenon becomes essentially low-dimensional, i.e. it does not arise for large $N$ (Fig.~2a).  
Withal, the solitary angle 
$\eta_{sol}=2\eta^{\ast}+\pi$ is not equal $\pi$ any more, contrary to the case $\alpha=\beta=0$. 
As $\e$ crosses $\e_{cr}$, $\eta_{sol}$ splits up from 0 and monotonically
increases/decreases with $\e$, approaching eventually the value $(1-2/N)\tan\beta$ as $\e\rightarrow\infty$, see Fig.~2b.  
Increase or decrease of $\eta_{sol}$ is determined by the sign of the difference $\tan\alpha-(1-4/N)\tan\beta$; if 
the difference equals 0, $\eta_{sol}=2(N-2)/(N-4)\tan\alpha+\pi$ for all $\e>\e_{cr}$. 

\begin{figure}[ht!]
%\centerline{\includegraphics[width=0.48\textwidth]{cohexch.pdf}}
\centerline{\includegraphics[width=0.5\textwidth]{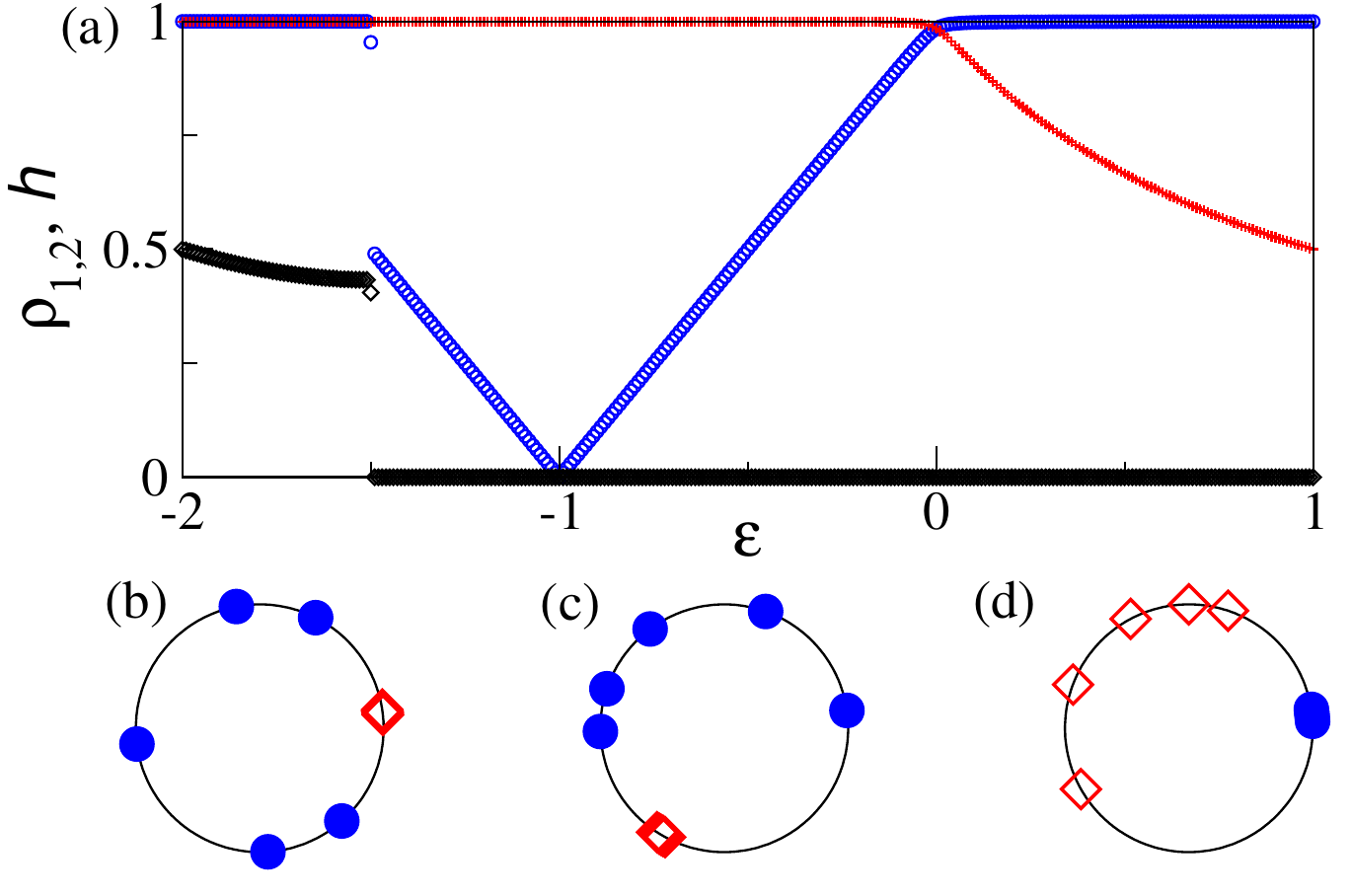}}
\caption{(Color online) (a) An example of coherence exchange for $N_{1,2}=5$, $\alpha=2\pi/3$, 
$\beta=0$. Here $\rho_{1,2}$ and  $h$ are shown by blue circles, red pluses, and black diamonds, respectively.
For $\e<\e_{cr}=-1.5$ all oscillators are synchronized. 
For $\e>\e_{cr}$ the repulsive units form a fuzzy cluster, $\rho_2\lesssim 1$, while the attractive oscillators first 
desynchronize, so that $\rho\approx 0$ for $\e=-1$  and then synchronize again. When $\e$ becomes positive, 
the attractive oscillators form a fuzzy cluster, while the repulsive desynchronize.
Initial phases for both groups are placed on two arcs of arbitrary length, shifted by $\pi$.
Panels  (b,c,d) are snapshots for $\e=-1$, $\e=-0.5$, and $\e=0.5$. 
}
\label{cohexch}
\end{figure}

We conclude that the desynchronization transitions at $\e=\e_{cr}$ immediately yields  the solitary state only if $\alpha=\beta$. Otherwise, there is always a gap between the full synchronization and the solitary behavior. Solitary states stabilize later at some $\e_{sol}^{-}>\e_{cr}$, and they lose their stability at $\e_{sol}^{+}>\e_{sol}^{-}$. 
The bifurcation values $\e_{sol}^{-}$ and $\e_{sol}^{+}$ depend on $N$, $\alpha$, $\beta$, $p$, and can be obtained from Eq.~(\ref{eqlam103}). 
For $\alpha\neq\beta$ and $N$ large enough, the solitary behavior does not appear and the transition can occur 
via the mechanism of \textit{coherency exchange}, illustrated in Fig.~\ref{cohexch} for $N_{1,2}=5$, $\alpha=2\pi/3$, and 
$\beta=0$.   As follows from Eq.~(\ref{hforce}), the condition $h=0$ implies that $\rho_a=0$ for $\e=-1$; 
the straight lines for $-1.5<\e<0$ and the hyperbolic curve for $\e>0$ are also 
explained by  Eq.~(\ref{hforce}).

\begin{figure}[ht!]
%\centerline{\includegraphics[width=0.48\textwidth]{vdpros.pdf}}
\centerline{\includegraphics[width=0.5\textwidth]{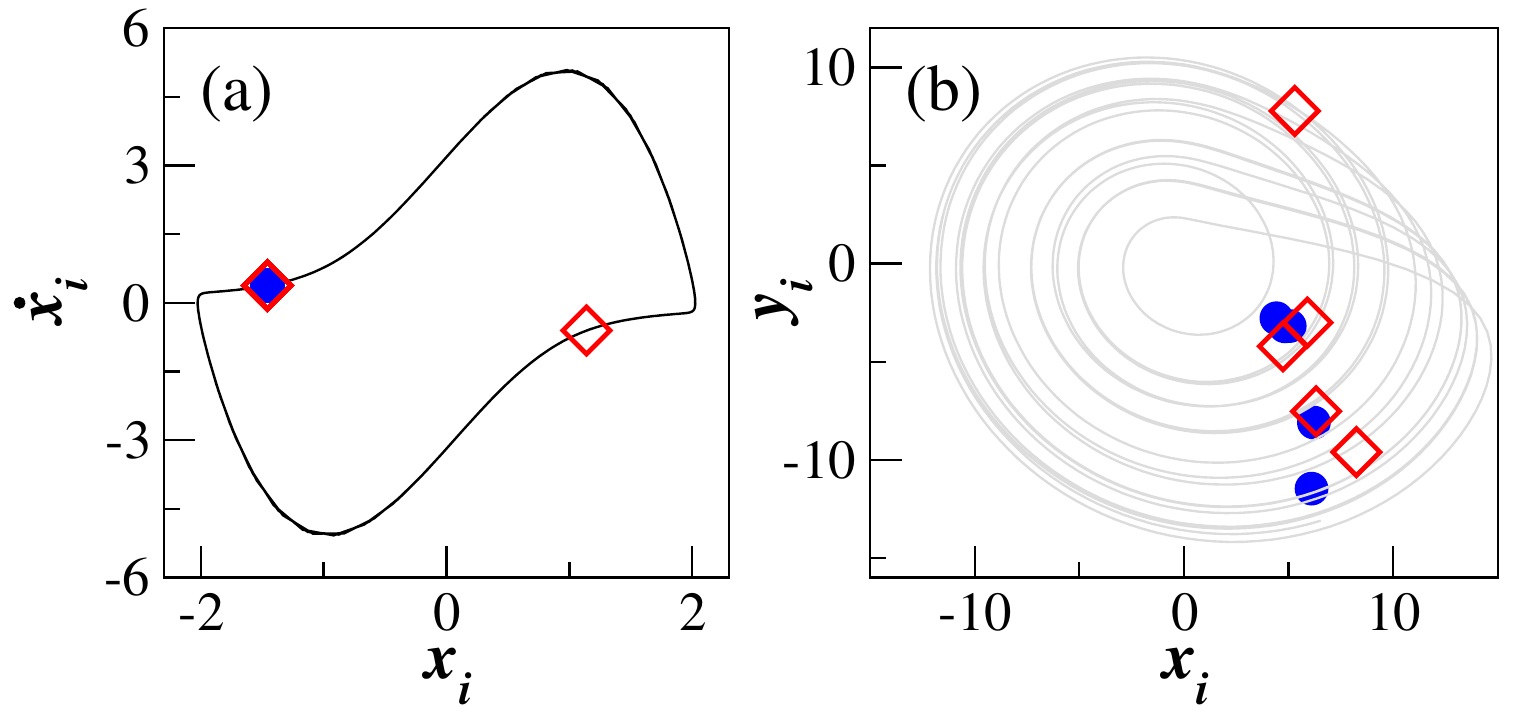}}
\caption{(Color online)  Solitary state in the ensemble of  van der Pol
oscillators (a) and of chaotic R\"ossler oscillators (b).
 Blue circles and red diamonds show the elements of the attractive and the repulsive 
 group, respectively.
Gray line shows the limit cycle of a single oscillator (a) and 
projection of the strange attractor of an an autonomous R\"ossler system (b). 
In (a) all oscillators except one are in the cluster at $x_j\approx-1$; the solitary element is at 
$x_j\approx 1$, i.e. nearly in antiphase.  
In (b) all oscillators except for one repulsive element group with approximately the same phase.
}
\label{vdpros}
\end{figure}

To test if the solitary states appear in more general networks, we analyzed 
a three-group Kuramoto model
\begin{equation}
 \dot\theta^{\sigma}_i = \sum_{\sigma'=1}^{3}\frac{K_{\sigma'}}
{N}\sum_{j=1}^{N/3}\sin[\theta^{\sigma'}_j-\theta^{\sigma}_i-(\sigma'-1)\frac{2\pi}{3}]  
\label{eqmod3}
\end{equation}
where $\sigma,\sigma'=1,2,3$, $K_{1,2}=1$, $K_3=1+\e$. Here the first group is attractive, and two others are 
repulsive, quantified  by  phase shifts $\mp 2\pi/3$. If $\e>0$, 
the repulsive action of the third group is stronger. Then, a solitary state is born at $\e_{cr}=0$, its stability region 
has the same shape as in the two-group $\alpha=\beta$ case: $0<\e<6(N-6)^{-1}$. 
As it is expected, solitary oscillator belongs to the third group, it splits up from all others remaining fully 
synchronized by the angle
$\eta_{sol}=\arctan\frac{3\sqrt{3}(1+\e)}{N\e}
+\arcsin\frac{\sqrt{3}(6(1+\e)-N\e}{2\sqrt{27(1+\e)^{2}+N^{2}\e^{2}}}+\pi$. 
The analysis of $M$-group models, $M\ge 4$, remains a subject of future studies; the preliminary analysis 
for $M=4$ and  
$\alpha_{\sigma}=(\sigma'-1)\frac{\pi}{2}\;,\; \sigma'=1,\ldots,4$
does not reveal the solitary states.
To illustrate that the demonstrated effect is not restricted to sine-coupled phase oscillators,
we performed numerical analysis of two more realistic models. 
First, we simulated globally coupled  van der Pol 
oscillators (cf. \cite{VazMartins-Toral-11}), for the case $N_{1,2}=20$:
\begin{equation*}
 \ddot x_j-3(1-x_j^2)\dot x_j +x_j= K_1 (X_1-\dot x_j)-K_2 (X_2-\dot x_j) \;,
\label{vdp}
\end{equation*}
where $X_1=N^{-1}_1 \sum_{k=1}^{N_1} \dot x_k$, 
$X_2=N_2^{-1}\sum_{k=N_1+1}^{N} \dot x_k$. 
Figure~\ref{vdpros}a exhibits a solitary state for $K_1=0.1$, 
$K_2=0.105$. 
Next, we consider attractively-repulsively coupled R\"ossler oscillators,  $N_{1,2}=5$,
\begin{equation*}
\begin{aligned}
 \dot x_j &= -y_j -z_j+K_1 (X_1 - x_j)-K_2 (X_2-x_j)\;,   \\
 \dot y_j &=  x+ 0.15 y +K_1 (Y_1 - y_j)-K_2 (Y_2-y_j) \;, \\
 \dot z_j &= 0.4 +z_j(x_j-8.5)\;,
\end{aligned}
\label{ros}
\end{equation*}
where  $X_1=N_1^{-1}\sum_{k=1}^{N_1} x_k$, $X_2=N_2^{-1}\sum_{k=N_1+1}^{N} x_k$,
 $Y_1=N_1^{-1}\sum_{k=1}^{N_1} y_k$, $Y_2=N_2^{-1}\sum_{k=N_1+1}^{N} y_k$.
Since the systems are chaotic we can expect only some quantitative correspondence with our
theory. Indeed, we observe a narrow domain with (approximately) solitary states, see 
Fig.~\ref{vdpros}b, where $K_{1,2}=0.05$.

% \begin{equation}
% \begin{array}{c}
%  \dot\vp_i= \w+\frac{K_{11}}{N_1}\sum_
% +\frac{K_{12}}{N_2}\sum_{j=1}^{N_2}\sin(\psi_j-\vp_i),   \\[3ex]
%  \dot\psi_i=\w+\frac{K_{21}}{N_1}\sum_{j=1}^{N_1}\sin(\vp_j-\psi_i)  
% +\frac{K_{22}}{N_2}\sum_{j=1}^{N_2}\sin(\psi_j-\psi_i),   
% \end{array}
% \label{eqmod}
% \end{equation}
% 
% 
% 

In conclusion, we have identified a novel scenario for the coherence-incoherence transition in networks 
of globally coupled identical oscillators with attractive and repulsive interactions.  
The transition occurs via solitary state at the edge of synchrony.
The phenomenon arises when attraction and repulsion act in antiphase 
and diminishes and becomes low-dimensional when this condition is not exact. 
In the desynchronized state the system is highly multistable;  in particular it exhibits 
fuzzy clustering.  
Finally, we have found solitary states for more realistic oscillatory networks with both periodic and chaotic 
local dynamics. 
This indicates a general,
probably universal desynchronization mechanism in networks of very different nature, 
due to attractive and repulsive interactions.

We are greatly acknowledge fruitful and illuminating discussions with A. Pikovsky, E. Sch\"oll, and T. Girnyk. 
Yu. M. acknowledges financial support from the Merkator-Stiftung, Germany.

%\bibliography{excinh}
%Merlin.mbs v4.21 2009-07-09.
%

\end{document}